\documentclass[a4paper,11pt]{article}
\input{colordvi}
\usepackage{epsf}
\usepackage{graphicx}        % standard LaTeX graphics tool
\def\ll{\label}
\def\re{\ref}
\def\c{\cite}

\def\r1{(\ref{$1})}

\def\th{\theta}
\def\ba{\begin{array}{c}}

\def\ea{\end{array}}

\def\si{\sigma}

\def\De{\Delta}
\def\de{\delta}

\def\ha{{1\over 2}}

\def\l{\left}
\def\l({\left(}
\def\r){\right)}
\def\r{\right}

\def\la{\lambda}
\def\al{\alpha}

\def\be{\begin{equation}}
\def\bc{\begin{center}}
\def\ec{\end{center}}
\def\bit{\begin{itemize}}
\def\eit{\end{itemize}}
\def\ee{\end{equation}}
\def\ed{\end{document}}
\def\bea{\begin{eqnarray}}
\def\eea{\end{eqnarray}}
\def\efr{\end{flushright}}

\begin{document}
\title{
Quantum and classical integrable sine-Gordon model with  defect
}
%\vskip 1cm

\author{
Ismagil Habibullin\footnote{e-mail: habibullin\_i@mail.rb.ru,(On
leave from Ufa Institute of Mathematics, Russian Academy of
Science, Chernyshevskii Str. , 112, Ufa, 450077, Russia)},
\\Department of Mathematics, Faculty of
Science,
 \\Bilkent University, 06800, Ankara, Turkey \\
Anjan Kundu \footnote {Corresponding Author:
  email: anjan.kundu@saha.ac.in, Phone:+91-33-2337-5346, FAX:
+91-33-2337-4637} \\
  Saha Institute of Nuclear Physics,\\
 Theory Group \& Centre for Applied Mathematics and Computer Science, \\
 1/AF Bidhan Nagar, Calcutta 700 064, India.
 }
\maketitle
\vskip 1 cm

\begin{abstract}

Defects  which are predominant in a  realistic model, usually spoil
its integrability or solvability. We on the other hand
 show the exact integrability of a known  sine-Gordon
field model with a defect (DSG), at the classical as well as at the quantum level
based on the Yang-Baxter equation. We find the associated classical and
quantum
R-matrices and the underlying q-algebraic structures, analyzing the exact
lattice regularized model. We derive algorithmically all higher conserved
quantities $C_n, \ n=1,2,\ldots$ of this integrable DSG model, focusing explicitly on the
contribution of the defect point      to each $C_n$. The bridging condition
across the defect, defined through  the B\"acklund transformation is found to
induce creation or  annihilation of a soliton by the defect
point or its preservation with a phase shift.
\end{abstract}
%int system
, 11.15.Tk
%nonperturbative technique
, 02.20.Uw
%q-group
, 11.10.Lm
%nonlinear nonlocal FT
, 72.10 Fk
%with defect and impurity

\medskip


\section{Introduction}
Systems with defects and impurities are prevalent in nature.
Many theoretical  studies are  dedicated to  various  models with
defects starting from classical and semiclassical
 to quantum as well as statistical models
 \cite{defects}, with several of them devoted  exclusively to the
 sine-Gordon  (SG)  model with  defect (DSG) \c{dsg}  or inhomogeneity \c{msgi},
 which  have  enhanced
physical importance \c{msgni1,msgni2}.
 A specific form of  DSG  model  with a
 defect  at a single  point $x=0 $ :
           \be u^\pm_{tt}-u^\pm_{xx}+ {\rm sin} u^\pm=0, \ \ \mbox{ for} \
u^+=u(x \geq 0,t), \ u^-= u(x \leq 0,t)), \ll{dsg}\ee
exhibiting intriguing  properties close to  integrable systems was investigated
in a series of papers
 \c{dsg,qdsg}.
These investigations
 aimed
to find  out  mainly the  additional contribution of the defect point to the conserved
Hamiltonian and momentum of the system,
 using the
  Lagrangian formalism and  the effect of the defect point  on the
  soliton solution using the scattering theory.    Though
the studies were  concentrated
 basically on the  classical aspects of this  model,
some semiclassical and  quantum arguments were also put forward
\c{qdsg}. The  important central  idea  of this approach
  is the existence of an auto B\"acklund transformation (BT)
 frozen at the defect
point $x=0 $, relating  two solutions $u^\pm $ of the SG equation
along the
positive and negative semi-axis \c{dsg,qdsg}.

Our aim here is to establish the suspected integrability of
the above DSG \c{dsg}, by showing the existence of
 infinite set of its conserved
quantities and finding them explicitly. The idea is to  adopt
the monodromy matrix approach  expressed through matrix
Riccati equation \c{FT}, a true signature  of the
 integrable systems \c{solit}, and  couple
it with the
important  concept  of extending the domain of defect fields $u^\pm $
through BT
\c{semihab}. This approach yields apart from finding out systematically the
defect-contribution for all higher conserved quantities,
 an intriguing possibility of creation or annihilation of
soliton by the defect point.

More significantly,
exploiting   the ancestor model approach of \c{kunprl}
we find an exact  lattice regularized version of the  DSG model by using
the realizations of the underlying algebra. This
allows to solve
the long awaiting problem of establishing the
 complete integrability of this   model, by
finding  the  classical and the quantum R-matrix solutions and
 showing the exact solvability  of the classical and quantum Yang-Baxter
equations (YBE).
The  exact algebraic Bethe ansatz solution   can also be formulated  for the
 quantum DSG
model, though its explicit resolution needs further study.

\section{Bridging condition and Lax pair for the SG model with defect}

We focus  on the central idea in DSG model (\ref{dsg}) of gluing
its fields $u^\pm $ across the defect point through the  BT as
 \be
u_x^+(x,0)=u_t^-(x,0)+p(x)+q(x), \qquad
u_t^+(x,0)=u_x^-(x,0)+p(x)-q(x) \ll{btgx}\ee where \be
p(x)=a\sin\frac{ u^+(x,0)+u^-(x,0)}{2},  \ \ q=a^{-1}\sin\frac{
 u^+(x,0)-u^-(x,0)}{2}. \ll{pqgx}\ee
with parameter $ a$ signifying the intensity of the defect.
However, we  would like to stress  on an important conceptual
difference in the  role of BT (\ref{btgx}, \ref{pqgx}) played in
the present approach and that in the previous studies \c{dsg,qdsg}, where
the above BT was considered to be {\it frozen}  at the defect
point $x=0$, and hence  playing no role  at any other point: $x
\neq 0 $. Therefore,   since  the solutions $u^\pm (x) $ can not
be related  through BT at other points along the axis,
% except at the centre,
 soliton number remains   unchanged while moving across the
defect point \c{qdsg}. We on the other hand implement  here the idea of
 \c{semihab} used in  the semi-axis SG
model, where the  domain  of
the field  is extended
  with the application of  BT. Therefore
in place of a frozen BT
we use (\ref{btgx}) effectively   at
all points of the axis including the defect  in the following sense.

Define a  solution of the SG equation
 with rapidly decreasing initial data
\[ u(x,0)= \left \{
\begin{array}{ll} u^-(x,0)
  & \mbox{if}\quad x\leq0 \\
 u^+(x,0)     & \mbox{if}\quad x\geq0  \end{array}
 \right. \ \
\mbox {and} \ \
u_t(x,0)= \left \{
\begin{array}{ll} u_t^-(x,0)
  & \mbox{if}\quad x\leq0 \\
 u_t^+(x,0)     & \mbox{if}\quad x\geq0  \end{array}
 \right. \]
satisfying the gluing conditions (\ref{btgx}) and
having the limits $\lim_{|x|\rightarrow\infty}u(x,0)=0$.
This field  solution of the SG equation
allows to extend the pair of functions  $u^-(x,0)$, $u_t^-(x,0)$
 smoothly from the left half-line $x\leq0$ onto the whole line using the BT
(\ref{btgx})
with a  limiting value  at the positive infinity  $x\rightarrow+\infty$ :
$u^-(x,0)\rightarrow2\pi m_-$ with an
integer $m_-$.
Following \c{semihab} one can prove also  the  existence
 and uniqueness of such an extension.
%the followingIt can be proved (see for more details \cite{BT2}) that the smooth
 Similarly one can prolong the other pair of functions
$u^+(x,0)$, $u_t^+(x,0)$ from the right half-line to the whole
line by means of the same BT and get
$u^+(x,0)\rightarrow2\pi n_+$, at $x\rightarrow-\infty$, $n_+$ being  another
integer. Now one has two potentials $u^+(x,0)$, $u_t^+(x,0)$ and
$u^-(x,0)$, $u_t^-(x,0)$ related to each other by the BT.
 If the function $u(x,t)$ satisfies the  DSG
equation then the functions $u^+(x,t)$ and $u^-(x,t)$ solve the
usual SG equation. However in the context of the DSG which is the
focus model here, such extensions can be considered to be virtual and
used for mathematical manipulations, while  the physically
observable fields are only $u^- $ in the domain $ x<0$ and
similarly $u^+ $ in  $ x>0$.   Therefore any solution $u^- $
moving from the left along the axis $x<0 $ would be transformed
after crossing the defect at $x=0 $  to a solution $u^+ $ in the
region $x>0 $, determined through the relations (\ref{btgx}).
Therefore, as we see below, it opens up the possibility of
creation or annihilation of soliton by the defect point, which was
prohibited in earlier studies due to consideration of a frozen BT
relation \c{qdsg}. Apart from these solutions,  a single soliton
  suffering  a phase shift,  while propagating across the defect
point, as found earlier  \c{dsg,qdsg}, seems also to be present.
Interestingly, the BT expressed through scalar relations
(\ref{btgx})   can be incorporated more efficiently into the
machinery of integrable systems by representing it as a { gauge
transformation} relating the Lax pairs of the DSG: \be
U(u^+)=F^0U(u^-)(F^0)^{-1} +F^0_x(F^0)^{-1}, \ \
V(u^+)=F^0V(u^-)(F^0)^{-1} +F^0_t(F^0)^{-1} \ll{btgt} \ee where
$F^0(\xi,u^+, u^-) $ is the B\"acklund matrix (BM)
 \be
F^0(\xi,u^+, u^-) = e^{- {i \over 4}\sigma_3   u^-} M(\xi, a)
e^{ {i \over 4}\sigma_ 3  u^+}, \ \ M(\xi, a)=
 \left( \begin{array}{c}
\xi
 \qquad \ \ a
 \\
    \quad  -a
 \qquad \ \ \xi
          \end{array}   \right),  \ll{F0} \ee
involving both fields $u^\pm $
and bridging between them at all points, including the defect point  $ x=0$.
We can check directly  from the matrix BT relations
 (\ref{btgt}) that by inserting the explicit form of  SG
Lax operators \c{solit}:
\bea
 \ U =  \frac 1 {4i} \left( u_t \si _3 +k_1 \cos {\frac u 2} \si _2
+ k_0 \sin {\frac u 2} \si _1 \right),
\ \
 V= \frac 1 {4i} \left( u_x \si _3 +k_0  \cos {\frac u 2} \si _2
+ k_1 \sin {\frac u 2} \si _1 \right), \ll{UV}\eea
where $ k_0=   \xi +\frac {1} { \xi}, \ \
k_1 =   \xi -\frac {1} { \xi}, $ with    spectral parameter $ \ \xi  $,
and comparing the matrix elements, one can derive the scalar BT relations
(\ref{btgx}). It is also obvious from (\ref{btgt}) using
the flatness condition $U_t-V_x+[U,V]=0 $  that if $u^- $ is a solution  of
the SG equation, so is $u^+ $. Note also that
since the corresponding Jost solutions are
related by $\Phi(\xi,u^+)= F^0(\xi,u^+, u^-)\Phi(\xi, u^-)$,  the
 exact N-soliton solution
 may
change its number by one, after crossing the defect point,
 a  possibility lost for the frozen  BT
 \c{qdsg}.
\section{Conserved quantities for DSG
 model  }

For deriving the infinite set of conserved quantities,  an essential
property of an integrable system,  for the SG model with
a defect we combine the matrix Riccati equation
technique
  for the standard SG model \c{FT} with the
idea of bridging scattering matrices through BT \c{semihab}.
Therefore let us first describe briefly the technique
developed by Faddeev-Takhtajan for the SG model.

\subsection{Conserved quantities for SG model}
Define the  monodromy  or the  transition matrix as a solution to the associated
linear equation
\begin{equation}\label{L0}
\frac{dT}{dx}(x,y,\xi)=U(x,\xi)T(x,y,\xi),
\end{equation}
with the initial data $T(y,y,\xi)=1$. To expand the transition
matrix  in asymptotic power series as $|\xi|\rightarrow\infty$,
it is convenient  to gauge transform  the variable
 \be \ T \to \tilde
T(x,y,\xi)=\Omega^{-1}(x) T(x,y,\xi)\Omega(y)
\label{ch}\ee
  in
equation (\ref{L0}), with $\Omega(x)=e^{\frac{i }{4}
u(x)\sigma_3}$ and represent
% the variable $\tilde T(x,y,\xi)$
it  as a
product of  {\it amplitude} and a {\it phase}
 \be \tilde
T(x,y,\xi)=(1+W(x,\xi))\exp (Z(x,y,\xi)) (1+W(y,\xi))^{-1},
\ll{wzw0}\ee
 where $W(x,\xi)$ is an off-diagonal and $Z(x,y,\xi)$
a diagonal matrix, satisfying the condition $ \
Z(x,y,\xi)|_{x=y}=0.$  By a direct substitution
of the gauge transformed $\tilde T $ in the form (\ref{wzw0}) into the
equation  (\ref{L0}) one gets
\be
Z(x,y,\xi)= \frac{1}{4i}\int_y^x(\theta(x')\sigma_3+
(\xi\sigma_2-\frac{1}{\xi}\sigma_2e^{iu(x')\sigma_3})
W(x',\xi))dx',\label{z}\ee
where $\theta(x)=u_t(x,t)+u_x(x,t)$ and
$W(x,\xi)$ solves a matrix Riccati equation
\begin{equation}\label{ric}
\frac{dW}{dx}=\frac{1}{2i}\theta\sigma_3W+\frac{1}{4i}
\xi(\sigma_2-W\sigma_2W)-\frac{1}{4i\xi}
(\sigma_2e^{iu\sigma_3}-W\sigma_2e^{iu\sigma_3}W),
\end{equation}
 This nonlinear
equation
due to  very special form of the coefficients admits asymptotic integration at $\xi \to \infty $,
\begin{equation}\label{asym}
W(x,\xi)=\Sigma_{n=0}^{\infty}\frac{W_n(x)}{\xi^n},
\end{equation}
where $W_0=i\sigma_1$. Putting expansion (\ref{asym})  in (\ref{ric})  and
comparing the coefficients with different   powers of $\xi $ we get the recurrence relation
\begin{eqnarray}\label{recurr}
W_{n+1}(x)=2i\sigma_3\frac{dW_n(x)}{dx}-
\theta(x)W_n(x)+\frac{i}{2}\Sigma_{k=1}^{n}W_k(x)\sigma_1W_{n+1-k}(x)-\\
-\frac{i}{2}\Sigma_{k=0}^{n-1}W_k(x)\sigma_1e^{iu(x)\sigma_3}
W_{n-1-k}(x)-\frac{i}{2}\sigma_1e^{iu\sigma_3}\delta_{n,1},
 \,\mbox{for}\,n=0,1,... \nonumber
\end{eqnarray}
The corresponding expansion for
%\begin{equation}\label{Z}
$Z(x,y,\xi)=\frac{\xi(x-y)}{4i}\sigma_3+i\Sigma_{n=1}^{\infty}
\frac{Z_n(x,y)}{\xi^n},
$ yields from (\ref{z}):
%\end{equation}
\begin{equation}\label{Zn}
Z_n(x,y)=\frac{1}{4}\int_y^x\sigma_2(e^{iu(x')\sigma_3}W_{n-1}(x')
-W_{n+1}(x')dx',
\end{equation}
where matrices $W_n, \ Z_n$ are of the form
 \be W_n(x)=
%\left(\begin{array}{c} \quad0 \qquad \ \
 -\bar w_n(x) \sigma_ ++
 w_n(x)\sigma_ -
% \qquad \ \ 0         \end{array}   \right)
, \qquad
Z_n(x)=\frac 1 2(
%\left(\begin{array}{c}
  z_n(x)  (I+ \sigma_3) +
% \qquad \ \ 0 \\ \quad0 \qquad \ \
-\bar z_n(x) (I- \sigma_3))
%          \end{array}   \right)
\label{WZ} \ee and relations above can be written as the recursion
relations starting from $n \geq 1 $
%\begin{equation}\label{w0}

%\end{equation}
%and
\begin{eqnarray}\label{recurrwz}
w_{n+1}(x)=\frac{2}{i}\frac{dw_n(x)}{dx}-
\theta(x)w_n(x)+\frac{i}{2}\Sigma_{k=1}^{n}w_k(x)w_{n+1-k}(x)-\\
-\frac{i}{2}\Sigma_{k=0}^{n-1}w_k(x)e^{iu(x)}
w_{n-1-k}(x)-\frac{i}{2}e^{iu}\delta_{n,1},
  \nonumber\\ \mbox{and} \
 z_n(x,y)=\frac{i}{4}\int_y^x(w_{n+1}(x')-e^{-iu(x')}w_{n-1}(x'))dx'.
\end{eqnarray}
with $w_0=i $.
To derive finally the set of conserved quantities
we take the limit of the monodromy matrix
$ \tilde
T(x,y,\xi)_{x\rightarrow +\infty,y\rightarrow -\infty}=T(\xi)=e^{P(\xi)}+O(|\xi|^{-\infty})$
%  as $x\rightarrow +\infty$:\begin{equation}\label{lim}
%T(\xi)=e^{P(\xi)}+O(|\xi|^{-\infty}),
%\end{equation}
where
\begin{equation}\label{P}
P(\xi)= \frac 1 2(
%i\left(\begin{array}{c}
 p(\xi)   (I+ \sigma_3)   -\bar p(\xi)   (I- \sigma_3)), \
%\begin{equation}\label{plim}
p(\xi)=\lim_{x\rightarrow +\infty,y\rightarrow
-\infty}(\Sigma_{n=1}^{\infty}
\frac{z_n(x,y)}{\xi^n}-\frac{1}{4\xi}(x-y)).
\end{equation}
As shown  in \cite{FT}
  the  generating function of  the conserved
quantities :
%\begin{equation}\label{log}
$p(\xi)=\log a(\xi)=i\Sigma_{n=1}^{\infty}\frac{C_n}{\xi^n}, $ \
{at} \ $|\xi|\rightarrow\infty $.
%\end{equation}
is obtained by solving the recurrence equation (\ref{recurrwz}) as
\begin{equation}\label{i1}
C_1=-\frac{1}{4}\int_{-\infty}^{+\infty}(\frac{1}{2}
(u_t(x)+u_x(x))^2+(1-\cos u(x))dx
\end{equation}
and for arbitrary $n>1$
\begin{equation}\label{in}
C_n=\frac{i}{4}\int_{-\infty}^{+\infty}(w_{n+1}(x)-
e^{-iu(x)}w_{n-1}(x))dx.
\end{equation}
To derive the asymptotic expansion for $\xi\rightarrow 0$ it
suffices to use the involution \c{FT} $(\xi,\pi,u)\rightarrow
(-\xi^{-1},\pi,-u)$, with $\pi=u_t$, which leaves the Lax pair
invariant. As a result we get
%\begin{equation}\label{log-}
$ \ \log a(\xi)=i\Sigma_{n=1}^{\infty}C_{-n}\xi^n,\quad \mbox{as}\quad
\xi\rightarrow0,$
%\end{equation}
where $C_0=\frac{1}{2}\lim_{x\rightarrow+\infty}
u(x,t)$ and $C_{-n}(\pi,u)=(-1)^nC_n(\pi,-u),$ $n=1,2,...$,
giving in particular
$C_{-1}=-\frac{1}{4}\int_{-\infty}^{+\infty}(\frac{1}{2}
(u_t(x)-u_x(x))^2+(1-\cos u(x))dx
$. Therefore one can get the explicit form of
 the momentum $P$ and the Hamiltonian $H$ of the SG model as
\bea P&=&2(C_{-1}+C_1)= \int_{-\infty}^{\infty}P(u)dx, \
 P(u)=u_xu_t,  \nonumber \\
 H&=&2(C_{-1}-C_1)=\int_{-\infty}^{\infty}H(u)dx, \
 H(u)= \frac {1} {2}(u_x^2+u_t^2) +  (1- \cos  u ).
\ll{hp}\eea
%where we have scaled $u $ by a factor ${\alpha}$for notational
%convenience.
\subsection{Extension to DSG model }
We now extend the above result of the standard
SG model to the SG with a defect (DSG)
 showing  that the DSG equation  admits an infinite
set of conserved quantities indicating the integrability of this system. In
fact,  any conserved quantity \
$C_n=\int_{-\infty}^{+\infty}\rho _n(x,t)dx \ $
of the SG model can be transformed into a conserved quantity for
the DSG model by adding some extra term $D_n$, as the
 contribution from the defect, such that
\be C^d_{n}=\int_{-\infty}^{0}\rho _n(x,t)dx +D_n+
\int^{+\infty}_{0}\rho _n(x,t)dx . \ll{dCn}\ee
 Our aim   is to find an algorithm for evaluating the additional terms $D_n$,
 for which
 we  suitably modify the above  approach for the SG model \cite{FT} by
using (\ref{btgt}), a crucial relation in the DSG model.
%pp. 403-407)

In analogy with the SG  we  define  the monodromy
matrix of the DSG  as a solution to the
associated linear equation with a defect at the point $x=0$: \
%\begin{equation}\label{L}
$ \ \frac{dT}{dx}(x,y,\xi)=U(x,\xi)T(x,y,\xi), \quad
x\neq0,\quad y\neq0
\ $
%\end{equation}
with the initial data $T(y,y,\xi)=1$. At the point $x=0$ we
have the jumping condition
\begin{equation}\label{jump}
T(0+,y,\xi)=\frac{1}{\xi-ia}F^0_0(\xi)T(0-,y,\xi),\quad
y\neq0,
\end{equation}
where $F^0_0(\xi) $ is the crucial gluing
 operator (\ref{F0}) at the defect point taking naturally the form
 \begin{equation} F^0_0(\xi)=\Omega ^{-1}(0-)M(\xi,a)
%\left(\begin{array}{cc} \xi &a\\ -a& \xi \end{array} \right)
 \Omega (0+), \ \
\mbox {where} \ \  \Omega (0\pm) =\exp
({i \sigma_3u(0\pm)\over4})
 . \label{F00} \end{equation}
Similar to the SG case
 we gauge transform
$T \to \tilde T(x,y,\xi)$ as in (\ref{ch})
%=\Omega^{-1}(x) T(x,y,\xi)\Omega(y)$  with $\Omega(x)=e^{\frac{i }{4}
%u(x)\sigma_3}$
and represent it as in (\ref{wzw0}),
% the form
%\be \tilde T(x,y,\xi)=(1+W(x,\xi)\exp (Z(x,y,\xi))
%(1+W(y,\xi))^{-1}, \label{wzw}\ee
% where $W(x,\xi)$ is off-diagonal and
%$Z(x,y,\xi)$ is a diagonal matrix, satisfying the condition $ \
%Z(x,y,\xi)|_{x=y<0}=0.$
where   $W$ solves a
Riccati type equation (\ref{ric}) and $Z$ is found explicitly in
terms of $W$ as in (\ref{z}).
%\cite {FadTah}.
%formulas (4.78)-(4.83)).

For finding   the   conserved quantities $C^d_n $, though we use
again the same   expansion at $\xi \to \infty  $:
%\begin{equation}\label{Z-}
$ \ Z(x,y,\xi)=\frac{\xi(x-y)}{4i}\sigma_3+
i\sum_{n=1}^{\infty}Z_{n}(x,y)\xi^{-n}, \ $
%\end{equation}
 the elements of the diagonal matrices
$ \ Z_n(+\infty,-\infty)=Z_n(+\infty,0+)+Z_n(0+,-\infty)=
Z_n(+\infty,0+)+\frac 1 i D_n+Z_n(0-,-\infty), \ $
 have now the contribution from the
defect point: $-i D_n=Z_n(0+,-\infty)-Z_n(0-,-\infty) $.
 Therefore the general form of
the set of conserved quantities  may be given by $ C^d_n=
C_n^++C_n^0+C_n^-,
\ \ n=1,2,\ldots ,$ \ {
where} $C^d_n  =trace (\sigma_3
 Z_n(+\infty,-\infty) ),\ $ with  \be    C_n^+ =trace (\sigma_3
 Z_n(+\infty,0+) ),  C_n^- =trace (\sigma_3
 Z_n(0-,-\infty) ), \  C_n^0 =-i trace (\sigma_3
 D_n). \ll{Cnd} \ee
 Following therefore the above approach \cite{FT}
with our extension, in place of (\ref{Zn}) we arrive at
\be Z_n(+\infty,-\infty)=\frac{1}{4}\int_{-\infty}^0\sigma_2
(e^{iu\sigma_3}W_{n-1}(x)-W_{n+1}(x))dx+\frac 1 i D_n+\frac{1}{4}\int_0^{-\infty}\sigma_2
(e^{iu\sigma_3}W_{n-1}(x)-W_{n+1}(x))dx
\ll{Zn0}\ee where
$W_n=-w_n^{*} \sigma_++ w_n \sigma_-$
 are the known solution of the Riccati equation (\ref{recurrwz}).

For  deriving the defect contribution
$D_n, n=1,2,\ldots  $ explicitly,
introduce the limiting monodromy matrix
\be T_-(x,\xi)=\lim_{y\rightarrow
-\infty}T(x,y,\xi)E_-(y,\xi), \ \ \mbox{ where} \ \
E_-(x,\xi)=e^{\frac i 2 \pi n \sigma_3} E(x,\xi), \ E(x,\xi)=\frac{1}{\sqrt2}\left(
\begin{array}{cc} 1 &i\\ i& 1 \end{array} \right)
e ^{\frac{1}{4i}(\xi-\xi^{-1})\sigma_3x}
,\ll{T-}\ee with $\lim_{y\rightarrow
-\infty}u(x,t)= 2\pi n $
Using $\Omega(-\infty)=1 $ and the jumping condition (\ref{jump}) we get
\begin{equation}\label{jump2}
\tilde T_-(0+,\xi)=\tilde F_0(\xi)\tilde T_-(0-,\xi), \ \ \mbox {where} \ \
 \tilde T_-(x,\xi)=\Omega^{-1}(x)T_-(x)
\end{equation}
 and
\bea \tilde
F_0(\xi)&=&\frac{1}{\xi-ia}\Omega^{-1}(0+)F^0_0(\xi)\Omega(0-)=
e_+^{-\sigma_3} \left(
\begin{array}{cc} \xi &a\\ -a& \xi \end{array} \right)   e_+^{\sigma_3}
\nonumber \\
&=&\frac{\xi+H}{\xi-ia}, \ \ \mbox{where} \
H=a\left(
\begin{array}{cc} 0 &A^{-1}\\ -A & 0 \end{array} \right), \quad
A=e_+^{2}, \ \  e_\pm= e^{ {i \over 4} (u^+(0)\pm  u^-(0))}
%=\exp ( \frac i 2 {u^+(0)+ u^-(0)}).
\label{tildeF}\eea
%where $e^+ $ is as defined in (\ref{rfd}).
 Using  (\ref{wzw0}) and (\ref{T-}) it follows  from (\ref{jump2}) that
$(1+W(0+,\xi))\exp (Z(0+,-\infty,\xi))=\tilde
F_0(\xi)(1+W(0-,\xi))\exp (Z(0-,-\infty,\xi))$ or readjusting,
\begin{equation}\label{jump3}
(1+W(0+,\xi))\exp D(\xi)=\tilde F_0(\xi)(1+W(0-,\xi))
\end{equation}
where $D(\xi)=Z(0+,-\infty,\xi)-Z(0-,-\infty,\xi)$
 is responsible for
  generating the addition     to the
 conserved quantities due to the
defect.

 Note that in equation
(\ref{jump3}) the two unknown quantities
 $D(\xi)$ and $W(0+,\xi)$ should be determined through two
 other  known quantities $\tilde F_0(\xi)$ and $W(0-,\xi)$, where
$\tilde F_0(\xi)$ is given explicitly as (\ref{tildeF}) and
 $W(0-,\xi)$ is a solution of the known   Riccati
type equation for the SG model
%\cite{FadTah}
%, equation (4.78))
 for the
half-line $x\in (-\infty,0)$.
%It is easy to check that the jump matrix $\tilde F$ can be written as follows
%$$\tilde F(\xi)=\frac{\xi+H}{\xi-ia},$$
%where $$H=a\left(
%\begin{array}{cc} 0 &A^{-1}\\ -A & 0 \end{array} \right), \quad
%A=\exp (i\frac{u+\bar u}{2}). $$
%For finding the conserved quantities with defect our primary
%concern is to evaluate $D(\xi)$ from (\ref{jump3}),
 For solving   $D(\xi)$     we consider
expansion for large values of  $\xi \to \infty  $:
\be \exp D(\xi)=\exp(\sum_{n=1}^{\infty}D_n\xi^{-n})=
1+\sum_{n=1}^{\infty}\tilde D_n\xi^{-n}, \ \ \mbox
{and}  \ 1+W(x,\xi)=\sum_{n=0}^{\infty}W_n(x) \xi^{-n} \ll{DW-xi}\ee
 and similarly  for vanishing values of
$\xi \to 0  $:
 \be \exp D(\xi)=\exp(\sum_{n=1}^{\infty}D_{-n}\xi^{n})=
1+\sum_{n=1}^{\infty}\tilde D_{-n}\xi^{n}, \ \ \mbox  {and}
  \ \ 1+W(x,\xi)=\sum_{n=0}^{\infty}W_{-n}(x) \xi^{n}.
\ll{DW+xi}\ee
Let us  evaluate first the  case with large $\xi $
which yields from
(\ref{jump3}) using
(\ref{DW-xi})
the  equation
\begin{equation}\label{jump4}
(\sum_{n=0}^{\infty}W_n(0+)\xi^{-n})(\xi-ia)
(1+\sum_{n=1}^{\infty}\tilde
D_n\xi^{-n})=(\xi+H)\sum_{n=0}^{\infty}W_n(0-)\xi^{-n},
\end{equation}
where $W_0(x)=1+i\sigma_1$.
 Gathering
coefficients before different powers of $\xi$ in the matrix equation
(\ref{jump4})
 one gets a recurrent
procedure  for solving  the off-diagonal matrices $W_n(0+)$
 and
diagonal matrices $D_n$ from the knowledge of $W_n(0-)$ and $ H $:
\begin{eqnarray}
  \xi &:& W_0(0+)=W_0(0-), \ll{xi}\\
  \xi^0 &:& W_1(0+)-iaW_0(0+)+W_0(0+)\tilde D_1=
  W_1(0-)+HW_0(0-), \ll{xi0}\\
  \xi^{-1} &:& W_2(0+)-iaW_1(0+) +(W_1(0+)-iaW_0(0+))\tilde D_1
  +W_0(0+)\tilde D_2 \nonumber \\& & \ \ =W_2(0-)+HW_1(0-),\ll{xi-}
\end{eqnarray}
and so on. For instance, the first nontrivial result is obtained from (\ref{xi0}): $\tilde
D_1=D_1=ia+iH\sigma_1$ yielding
 \be C^0_1=-i \ trace ( \sigma_3  D_1)=a(A+A^{-1})
= 2a \cos \frac {(u^+(0)+u^-(0))} {2},\ll{D10}\ee
 as a  contribution of the defect point to the conserved quantity $C^d_1 $.

For finding next $D_2$ use (\ref{xi-})  rewriting it as
$W_2(0+)-iaW_1(0+) +(W_1(0+)-ia(1+i\sigma_1))\tilde D_1
+(1+i\sigma_1)\tilde D_2=W_2(0-)+HW_1(0-)$. By taking the diagonal
part of this matrix equation one gets $\tilde
D_2=HW_1+ia\tilde D_1$, which using the relation
 $\tilde
D_2=D_2+\frac{1}{2}D_1^2$   yields
\be D_2=HW_1(0-)+iaD_1-\frac{1}{2}D_1^2,\ll{D2}\ee
 where $W_1(0-) $ is obtained by solving the
Riccati equation  as $w_1(x)=-i (p^-(x)+u^-_x(x)). $ Therefore
%$ W_n=\left(\begin{array}{cc} 0 &-w_n^{*}\\ w_n & 0 \end{array} \right)$
$C^0_2=-i \ trace (\sigma_3  D_2)$  is the    contribution of
  the defect point
to the conserved quantity $C^d_2 $ . In this recurrent way
we can find systematically  the contribution of the defect point at $ x=0$
 to all
higher conserved quantities  for this integrable DSG model.
Note   that one can also   explicitly determine from the above equations
 % (\ref{\xi}, \ref{\xi0}) that
\begin{eqnarray}\label{Wn+}
   & & W_0(0+)=W_0(0-)= 1+i\sigma_1 \nonumber\\
   & & W_1(0+)=W_1(0-)-a\sigma_1-i\sigma_1\tilde D_1+H=
W_1(0-)+\sigma_1 H \sigma_1+H
\end{eqnarray}
etc.
showing the effect  of the defect on
  the
monodromy matrix across the defect point.

Now we switch over to the complementary case $\xi \to 0  $ and
look for the conserved quantities $C^d_{-n} =trace (\sigma_3
 Z_{-n}(+\infty,-\infty) ) $ through the
expansion
\begin{equation}\label{Z+}
Z(x,y,\xi)=-\frac{(x-y)}{4i \xi}\sigma_3+
i\sum_{n=1}^{\infty}Z_{-n}(x,y)\xi^{n}.
\end{equation}
We  have to perform now similar expansion in the positive
 powers of $ \xi$ in all
the above formulas noticing the crucial symmetry of the monodromy matrix
\cite{FT}
 $ \ \hat T_-(x, -\frac 1 \xi; -u, p)=T_-(x,  \xi; u, p)  \ ,$
which is obvious from the  symmetry of the SG Lax operator
(\ref{UV}). It is crucial to note however that the unknown part of the
 monodromy matrix $W_{-n}(0+) $ across the defect point, as evident from
(\ref{Wn+}), depends on both $u^+,u^- $ and obviously the above symmetry is lost.
 Therefore we use  this symmetry  only for $ u^- \to -u^-$ (with same $ u^-_t$)
 without changing
the field $u^+ $ (but with $ u^-_t \to -u^-_t$, to preserve the canonical
structure),
 and expect the  consistent
 solution of  the
jump condition.  Therefore
in place of (\ref{jump2}) we get  the  condition
\begin{equation}\label{jump2+}
{\hat {\tilde T}}_-(0+, \xi)={\hat {\tilde F}_0}(  \xi) {\hat
{\tilde T}}_-(0-, \xi), \ \ \mbox {where} \ \
{\hat { \tilde T}}_-(0-,\xi; u^-, p^-)=\tilde
T_-(x,- \frac  1 \xi;-u^-, p^-)
\end{equation}
for the known solution of the Riccati equation   and
\bea {\hat {\tilde F}_0}(\xi)&=&
\tilde {F}_0(- \frac 1 \xi,  \frac 1 a; -u^-,u^+)
%\frac{1}{-( \frac 1 \xi+ \frac i a)}\Omega^{-1}(0+)F_0(\xi)\Omega(0-)=
\frac{1}{-( \frac 1 \xi+ \frac i a)}e_-^{-\sigma_3} \left(
\begin{array}{cc}-\frac 1 \xi & \frac 1 a\\ - \frac 1
 a& -\frac 1 \xi \end{array} \right)   e_- ^{\sigma_3}
\nonumber \\
&=&\frac{ \frac 1 \xi-\hat H}{\frac 1 \xi+ \frac i a}, \ \ \mbox{where} \
\hat H=\frac 1 a\left(
\begin{array}{cc} 0 &\hat {A}^{-1}\\ -\hat {A} & 0 \end{array} \right), \quad
\hat A=e_-^2=\exp ( \frac i 2  ({u^+(0)- u^-(0)})),
\label{hattildeF}\eea
Note that in (\ref{hattildeF}) we have made the transformation
 $\xi \to -\frac 1 \xi, a \to -\frac 1 a$ and $u^-(0) \to
-u^-(0),$ preserving  $ p^-(0) \to p^-(0)$  and $  u^+(0) \to
u^+ (0) $, which  demands
 also
$p^+_0 \to -p^+_0 $  for ensuring the corresponding quantum defect
matrix $F^d_0 $ (\ref{qdsgF}) to be a solution of the QYBE at the discrete level.
 This however does  not affect
(\ref{hattildeF}) obtained in the continuum.

Considering the above we obtain  the corresponding matrix
equations
\begin{equation}\label{jump4+}
(\sum_{n=0}^{\infty}\hat W_{-n}(0+)\xi^{n})(\frac 1 \xi+ \frac i a)
(1+\sum_{n=1}^{\infty}\hat {\tilde
D}_{-n}\xi^{n})=(\frac 1 \xi- \hat H)\sum_{n=0}^{\infty} \hat
W_{-n}(0-)\xi^{n}.
\end{equation}
Arguing in a similar way we get finally the required solutions
\bea \hat D_{-1}&=&-i( \frac 1 a + \hat H\sigma_1 ),\ll{D1+} \\
\hat D_{-2}&=&-\hat H \hat W_{-1}(0-)- \frac i a \hat  D_{-1}-
\frac{1}{2}\hat D_{-1}^2,\ll{D-2}\eea
 where $\hat W_{-1} $ is obtained from the
corresponding Riccati equation  through the
 solution  $\hat w_{-1}=-i (p^-(x)-u^-_x(x)). $
Note that the contribution
of the defect point to the conserved quantity  $C^d_{-1} $ is
\be C^0_{-1}= -i tr( \sigma_3  D_{-1})=-\frac 1 a(\hat A+\hat A^{-1})
= - \frac 2 a \cos \frac {(u^+_0-u^-_0)} {2},\ll{cD1+}\ee
while to $C^d_{-2} $ is  $C^0_{-2}=-i trace (\sigma_3 \hat D_{-2}).$
Therefore  we can derive the general form   for   conserved quantities
by using the simple symmetry
\[C^\pm_{-n}=(-1)^nC^\pm_{n}(p^\pm, -u^\pm  ), \ \  \hat D_{-n}=(-1)^n
D_{n}(-\frac 1 a,u^+(0),- u^-(0)), \]
from those obtained in  (\ref{Zn0}).
%, (see formulas (4.84), (4.86), (4.87))
%\subsection{Energy and momentum for the defect SG model}

Using  the conserved quantities derived above
we can  extend now the expressions  (\ref{hp}) for  the momentum and
 the  Hamiltonian  of the SG model to include
 the extra  contributions due to the defect point at $x=0 $:
\begin{equation}
\label{c1}
P^{(def)}=\int_{-\infty}^{0}P(u^-)dx+\int^{\infty}_{0} P(u^+)dx
-2a\cos \frac{ u^+(0)+u^-(0)}{2}+2a^{-1}\cos \frac{ u^+(0)-u^-(0)}{2}.
\end{equation} and
\begin{equation}\label{c2}
H^{(def)}=\int_{-\infty}^{0}H(u^-)dx+\int^{\infty}_{0} H(u^+) dx
-(2a\cos \frac{ u^+(0)+u^-(0)}{2}+2a^{-1}\cos \frac{ u^+(0)-u^-(0)}{2}),
\end{equation}
where the momentum  and Hamiltonian densities $P(u), \  H(u)$
%=u_xu_t
 are   given by their standard  expression  (\ref{hp}).
%and the  Hamiltonian as where $ H(u)
%= \frac {1} {2}(u_x^2+u_t^2) + (1- \cos u )
% $ is the standard Hamiltonian density
%of the SG model defined in  (\ref{hp}).

To convince ourselves  that (\ref{c1}, \ref{c2}) are  indeed
 conserved, we check it  by direct calculation.
For this we may  use an identity
$D_t(u_xu_t)=D_x(\frac{1}{2}(u_t^2+
u_x^2) + \cos u),$ which follows easily from the SG equation
$u_{tt}-u_{xx}=\sin u$.
 %$$D_t\int_{-\infty}^{\infty}u_xu_tdx=0$$
 Therefore noting that
 $u^\pm(x) $ together with their derivatives vanishes respectively
at $x=\pm \infty $, we get
\begin{equation}\label{Dc1}D_t P^{(def)}=\left(\frac{1}{2}((u^-_t)^2+
(u^-_x)^2) + \cos u^--\frac{1}{2}( (u^+_t)^2-
(u^+_x)^2) - \cos  u^+ +( u^+_t+u^-_t)p-( u^+_t-u^-_t)q\right)_{|x=0},
\end{equation}
  where
\be p=a\sin\frac{ u^++u^-}{2},  \ \ q=a^{-1}\sin\frac{ u^+-u^-}{2}. \ll{pq}\ee
Using now the  B\"acklund gluing condition at  $x=0 $
\be
u^+_x=u^-_t+p+q, \qquad  u^+_t=u^-_x+p-q \ll{bt}\ee
and consequently
\bea (u^+_x)^2&=&(u^-_t)^2+p^2+q^2+2u^-_tp+2u_tq+2pq, \nonumber \\
 (u^+_t)^2&=&(u^-_x)^2+p^2+q^2+2u^-_xp-2u_xq-2pq \ll{bt2}\eea
we can substitute  $ (u^+_x)^2$, $ (u^+_t)^2$, $ u^+_x$
and $ u^+_t$ through their expressions  above and apply the identity
$\cos u^+-\cos u^-=-2pq$ to derive from (\ref{Dc1})
 $D_t P^{(def)}=0$.

Turning now to  $H $ in (\ref{hp}) we
%$$D_t\int_{-\infty}^{\infty} H(u)dx=0$$
 use another identity
$D_t( H(u))=D_x(u_xu_t)$, which  follows  again from the SG
equation,
%$u_{tt}=u_{xx}-\sin u$.
and  we  show  similarly  that
$H^{(def)} $ is also a conserved quantity. Indeed we get
\begin{equation}\label{Dc2}D_t H^{(def)}=\left(u^-_t u^-_x - u^+_t  u^+_x +(
u^+_t+u^-_t)p+( u^+_t-u_t)q\right)_{|x=0}\end{equation}
 where $p $ and  $q $ are as defined in (\ref{pq}).
%$p=a\sin\frac{\bar u+u}{2}$ and $q=a^{-1}\sin\frac{\bar u-u}{2}$
Using again the BT (\ref{bt})  we can rewrite the
first part of (\ref {Dc2}) as
$u^-_t ( u^+_t -(p-q))- u^+_t ( u^-_t +(p+q)) $,
which clearly cancels with its second part to give zero,
 proving $ H^{(def)}$ to be   a conserved quantity.

\section {Soliton solution   in DSG with
 its possible creation \& annihilation}

We now find the relation between the scattering matrices linked to
two $ \pm $-regions and the intriguing contribution of the defect
point in creation or annihilation of the soliton. At the same time
using the BT  (\ref{btgt}) unfrozen at all points as explained
above we can find  soliton solutions showing explicitly their
creation, annihilation or preservation with a phase shift.

To clarify the procedure we introduce some definitions refining
that of (\ref{T-}), where we denote $T^{(\pm)} $ to indicate
monodromy matrix belonging to the fields $ u^\pm$, respectively.
Remind that the fields have the space asymptotics:
\bea
u^{\pm}\rightarrow2\pi m_{\pm} \quad\mbox{for} \quad
x\rightarrow+\infty ,\label{asyu+}\\
u^{\pm}\rightarrow2\pi n_{\pm} \quad\mbox{for} \quad
x\rightarrow-\infty \label{asyu-} \eea which provide the following
asymptotics for $T^{(\pm)}$

\be T^{(\pm)}_{-}(x,\xi)\rightarrow
e^{\frac{i\pi}{2}\sigma_3n_{\pm}}E(x,\xi)\quad\mbox{for} \quad
x\rightarrow-\infty \ll{T(-)}\ee and similarly
 \be T^{(\pm)}_{+}(x,\xi)\rightarrow
e^{\frac{i\pi}{2}\sigma_3m_{\pm}}E(x,\xi)\quad\mbox{for} \quad
x\rightarrow+\infty. \ll{T(+)}\ee

 We further relate the matrices involved using the
bridging condition as
 \be T^{(+)}(x,y,\xi)=F^0(x,
\xi)T^{(-)}(x,y,\xi)C(y,\xi),  \ll{T(pm)}\ee with  $ F^0(x, \xi)$
as in (\ref{F0}) and a matrix-valued function $ C(y,\xi)$ which
does not depend on $x$ but depends on $y$ and $\xi$. Using the
chain of relations (\ref{asyu+}-\ref{T(+)}), we can relate the
monodromy matrices  in the $\pm  $ region as \be
T_{\pm}^{(+)}(x,\xi)= F^0(x, \xi) T_{\pm}^{(-)}(x,\xi) \tilde
F_\pm^{-1}, \ \mbox{where} \ \tilde
F_\pm=diag(\xi+ia_{\pm},\xi-ia_{\pm}) \ll{T(+-)}\ee with
\be a_{+}=a(-1)^{m_++m_-} , \ \ a_{-}=a(-1)^{n_++n_-}. \ll{apm}\ee To get these
relations one has to compare asymptotics of the functions
$T^{(\pm)}_{\pm}$  at the infinities,
 choosing $C(x,\xi)$ through  $F^0(x, \xi) $ matrix.

Therefore from the definition of the scattering matrix
$S^{(\pm)}(\xi)=(T_{+}^{(\pm)}(x,\xi))^{-1}T_{-}^{(\pm)}(x,\xi) $
we  relate them as \be S^{(+)}(\xi)= \tilde F_+S^{(-)}(\xi)
 \tilde F_-^{-1}.
  \ll{S+-}\ee
Now from (\ref{S+-}) we get finally
 the relations between the scattering data
\[ s^+=\{a^+(\xi), b^+(\xi), \xi_1^+,\xi_2^+,...\xi_{n_+}^+;\gamma_1^+,
\gamma_2^+,...\gamma_{N_+}^+\} \ \mbox {and} \ s^-=\{a^-(\xi),
b^-(\xi), \xi_1^-,\xi_2^-,...\xi_{n_-}^-;\gamma_1^-,
\gamma_2^-,...\gamma_{N_-}^-\} \] as
 \be a^+(\xi)=a^-(\xi )
 \left ( \frac {\xi +i a_+} {\xi +i a_-}   \right ), \
 b^+(\xi))= b^-(\xi)  \left ( \frac {\xi +i a_+} {\xi -i a_-}   \right
)\ll{ab}\ee etc., where $ a_\pm $ as defined in (\ref{apm}) involve asymptotic $(m_{\pm},n_{\pm})$ for
the fields at space-infinities  and the defect intensity $a $.

There can be  three distinct possibilities \cite {semihab}:

1) $a_+=(-1)^{m_++m_-}a<0$, $a_-=(-1)^{n_++n_-}a>0, $ when soliton
number increases by $1 : $ $N_+=N _-+1$ (a soliton with $\xi_{N+}=
i a $ is created by the defect). We have $\xi_j^-=\xi_j^+,\quad
\gamma_j^-=\gamma_j^+$ for $j=1,2,...N_-,$ the set $S^+$ has an
extra eigenvalue $\xi^+_{N_+}$ compared with $S^-$, and
$a^+(\xi)=a^-(\xi)\frac{\xi-ia}{\xi+ia},$ $b^+(\xi)=b^-(\xi)$.

2) $a_+>0, a_-<0,$ when $N_+=N_--1$ and $\xi_j^-=\xi_j^+,\quad
\gamma_j^-=\gamma_j^+$ for $j=1,2,...n_+,$ the set $S^-$ has an
extra eigenvalue $\xi^+_{N_-}$ compared with $S^+$ and
$a^-(\xi)=a^+(\xi)\frac{\xi-ia}{\xi+ia}$, $ b^+(\xi)=b^-(\xi)$.

3)  $m_++m_-=n_++n_-$ ( mod $2$), when $N_-=N_+$. The sets $S^+$,
$S^-$ have the same number of eigenvalues and $\xi_j^-=\xi_j^+,
\quad\gamma_j^+=\frac{\xi_j+ia} {\xi_j-ia}\gamma_j^-\quad $ for
$j=1,2,...N_+,$ $a^+(\xi)=a^-(\xi),$
$b^+(\xi)=b^-(\xi)\frac{\xi+ia}{\xi-ia}.$

In  cases 1) and 2) there exists some extra {\it defect} soliton
with a very special behavior. Consider the case 1). If this
soliton moves to the right and originally is located on the left
half-line $x<0$ then it will appear as 2-soliton after defect i.e. a soliton
will be created for
$x>0$ (eee Fig 1a).
For $ u^-$ as 1-kink solution the boundary condition (BC) gives $ n_-=0, m_-=1$, 
while for  $ u^+$ as 2-kink solution it corresponds to  $ n_+=0,
m_+=2$, fullfilling the required condition that $ n_-+ n_+=0$  even, while
 $ m_-+ m_+=3$, odd. 
% For $m=n$, Soliton number remains same : $N_+=N_-$ \\({\it though
%soliton suffers
%phase-shift at  defect point }(see Fig
%3
 In 
case 2) we have a similar but opposite situation. A 2-soliton
 moving  to the right from $x<0 $
  will be converted into  1-soliton after the defect i.e. a soliton
can be annihilated by the defect point (see Fig 1b).
Here for $ u^-$ as 2-kink solution the BC can  give  $ n_-=-1, m_-=1$, 
while for  $ u^+$ as 1-kink solution it yields $ n_+=0,
m_+=1$, having  the required condition $ n_-+ n_+=-1$  odd , while
 $ m_-+ m_+=2$, even. 

We can derive such exact soliton solutions explicitly from the BT (\ref{btgt}). For
example inserting  for $u^- $,  1-kink solution in the Hirota form:
 $u^-=-2i\ln
\frac {f_+} {f_-},\ f_\pm= 1\pm f , \ f=e^{k_0 x+k_1 t +\phi_0} $, where
$k_0=\cosh \th , \ k_1= \sinh \th $
 we can extract from
  the BT a 2-kink solution for $u^+ $ in the form
 $u^+=-2i\ln
\frac {f_+} {f_-},\ f_\pm= 1\pm (f_1+f_2)+s(\frac 1 2(\th _1-\th_2))f_1f_2 ,
\ f_a=e^{k^{(\al )}_0 x+k^{(\al )}_1 t +\phi_\alpha } $,
with   scattering amplitude $s(t )= tanh ^2 t  $,
and with certain relations connecting the parameters $\th,\th _\alpha $ and
the defect parameter $a$.
For
 $ \la _2=-\la _1^*=\eta  e^{i\theta }$, one gets  the  kink-antikink
bound state (breather solution).

%\begin{figure}[t]
%\epsfxsize=0.96\textwidth
%\includegraphics[width=4.cm,height=3.1 cm]{sgd21s.eps}
%\caption{   Soliton
% ($s=\sin \frac u(x,t) 2 $) annihilation  by the
%defect point at $ x=0$ in integrable  DSG model}
%\end{center}
%\end{figure}

In  case 3) there is no  creation/annihilation of soliton
by the  {\it defect}. In this case soliton  passing through the
defect will suffer a phase shift of $\phi_{-+}= \log \frac {\eta +a}
{\eta -a} $, since the parameters $\gamma_j$ are changed. As also
shown in \c{dsg,qdsg},  if we insert 1-kink $u^-=-2i\ln \frac
{f_+} {f_-},\ f_\pm= 1\pm f_1 , \ f_1=e^{k_0 x+k_1 t +\phi_1} $
in BT we can again have 1-kink solution for $ u^+=-2i\ln \frac {\tilde f_+}
{\tilde f_-},\ \tilde f_\pm= 1\pm f_2 , \ f_2=e^{k_0 x+k_1 t +\phi_2}$, with a
phase shift given by $e^{\phi_1-\phi_2}=-\frac {\sinh d -\sinh \th
} {\cosh d +\cosh \th }$ where $k_0=2\cosh \th , \ k_1=2\sinh \th
, \ a+ \frac 1 a= 2 \cosh d, \ a- \frac 1 a= 2 \sinh d  $ (see Fig
1c).
Note that the BC for the kink solutions corresponds to  $ n_-= n_+=0 ,  m_-=
m_+=1$, giving $  m_-+
m_+= n_-+
n_+$ (mod 2), as predicted above. 
%\begin{figure}[t]
%\epsfxsize=0.96\textwidth
%\includegraphics[width=4.cm,height=3.1 cm]{sgd11s.eps}
%\caption{   Soliton
% ($s=\sin \frac u(x,t) 2 $) phase-shift   by the defect point at $ x=0$ }
%\end{center}
%\end{figure}
%\end{document}
%\end

\begin{figure}[t]

\includegraphics[width=4.cm,height=3.1 cm]{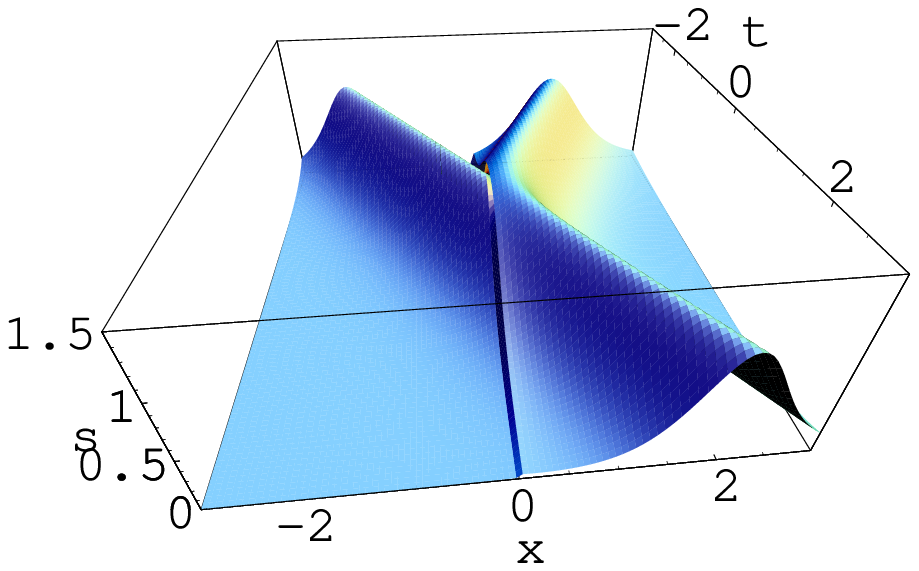}
 \ \
\includegraphics[width=4.cm,height=3.1 cm]{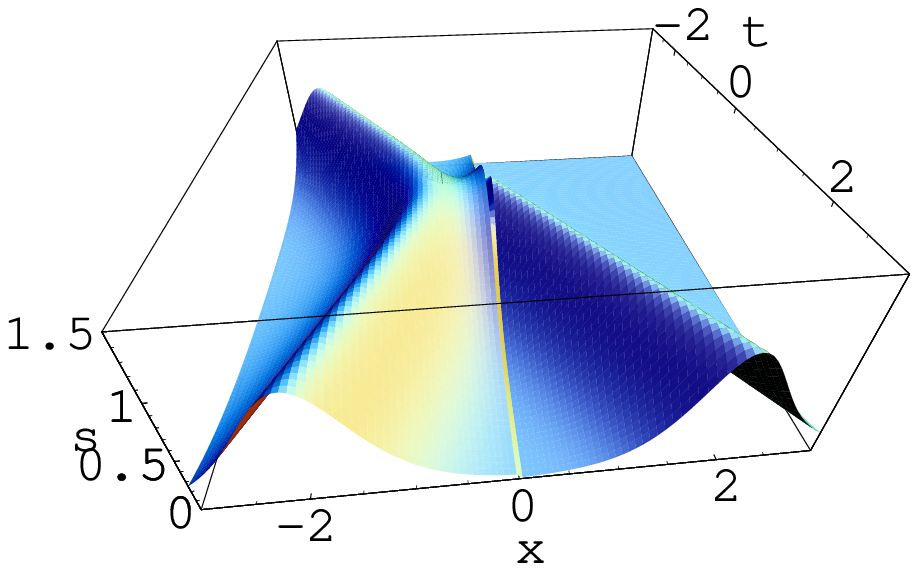}
\ \
\includegraphics[width=4.cm,height=3.1 cm]{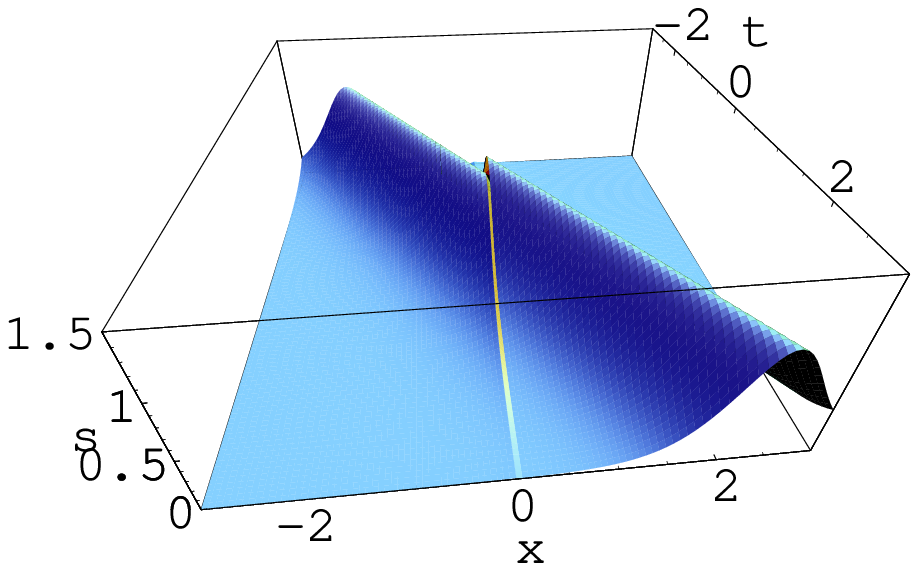}

\caption{   Soliton solutions  ($\sin \frac {u(x,t)}{ 2}
 $) for
DSG with a defect at $x=0$ showing a) creation, b) annihilation
and c) preservation with phase shift of soliton by the defect
point. }
%\end{center}
\end{figure}

\section{ Classical and Quantum integrability of DSG through Yang-Baxter
equation}
%%%%%%%%%%%%%%%%%
A semiclassical treatment of the DSG model through
factorizable S-matrix together with some possible quantum features are
presented   in \c{qdsg}. However for establishing   the  exact classical and
quantum  integrability,  it is
necessary to show the validity of the  Yang-Baxter equation  for this
 model
 both at the classical and the  quantum  level. Our aim  is to carry out
  this program by finding
the associated quantum and classical $R$-matrix  and the
lattice regularized Lax operators  for this system including the
defect point, as exact  solutions of the YBE. Subsequently we
formulate  the algebraic Bethe ansatz for the quantum DSG model.
 Our strategy is to follow closely   the approach  of
the standard quantum SG model  \c{qsg} in combination with the
ancestor model scheme of \c{kunprl}.

\subsection{Exact quantum integrability of lattice DSG model}
 We try to  construct
first an exact lattice  regularized   version of the quantum DSG model
 through a discrete  monodromy matrix
\be
T(\xi)= T^{N+}(\xi)
 F^d_0(\xi,u^+_0,  u^-_0) T^{N-}(\xi)
 \ll{qdsgT} \ee
where
\be
T^{N+}(\xi)=U^{+}_N(\xi, u^+_N) \cdots U^+_1(\xi, u^+_1)
, \ \   T^{N-}(\xi)=
U^-_{-1}(\xi, u^-_{-1})\cdots U^-_{-N}(\xi, u^-_{-N})
 \ll{qT+-} \ee
  with $U^\pm_j(\xi, u^\pm _j), \ j=\pm 1, \ldots \pm N $
% and $L^-_i(\xi, u_i), \ i=-1, \ldots -N $
being the discrete
 quantum Lax operator  of  the
lattice  SG model  defined  along both sides of  the defect, while
$ F^d_0(\xi,u^+_0, u^-_0)$ is the quantum Lax operator at the
defect point $j=0$. Recall that \c{qsg}  for  quantum
integrability  the monodromy matrix of the system (\ref{qdsgT})
must  satisfy  the global version of the  quantum YBE (QYBE) \be  R(\xi,\eta)T(\xi)\otimes
T(\eta)= (T(\eta)\otimes I) (T(\xi)\otimes
I)R(\xi,\eta),\ll{gqybe}\ee
which taking trace from both the sides yields evidently the relation $
[\tau(\xi), \tau(\eta) ]=0,$ where $\tau(\xi)=trace T(\xi)=\sum_n C_n\xi^n $,
giving finally the  quantum integrability condition
through the
commuting  set of conserved quantum operators as : $[ C_n, C_m ]=0 $.
 If we  ensure ultralocality condition, i.e.
 all constituent Lax operators $U^\pm_j, \ j \in [\pm
1, \pm N] $ and $ \ F^d_0, $
in (\ref{qdsgT}) situated at different lattice sites   mutually  commute,
then   it follows from (\ref{gqybe}),
 that  each of these local Lax operators  $U^\pm_j, \ F^d_0 $ must
also satisfy exactly the  local QYBE:
\be  R(\xi, \eta)L_j(\xi)\otimes L_j(\eta)=
(L_j(\eta)\otimes I) (L_j(\xi)\otimes I)R(\xi-\eta) \ll{lqybe}\ee
 with the same quantum $ R$-matrix, where $ L_j \equiv U^\pm_j $ at $ \ j \in [\pm
1, \pm N] $ and $L_0 \equiv F^d_0 $ at the defect point. The transition from local
QYBE (\ref{lqybe})  to the global one (\ref{gqybe}) is a reflection of the
coproduct
structure of the Hopf algebra property of the  underlying  quantum algebra
\c{qalg}.

It is known that the standard SG model satisfies (\ref{lqybe}) with trigonometric $R^{trig}
$-matrix \c{qsg}
\bea
& & R^{11}_{11}=R^{22}_{22}=a(\la , \al ), \ R^{11}_{22}=R^{22}_{11}=b(\la), \
R^{12}_{21}=R^{21}_{12}= c (\al ) \nonumber \\
\mbox {where } & & a(\la , \al )=\sin (\la +\al), \ b(\la )=\sin \la ,
\ c(\al )=\sin \al .
\ll{R}\eea
Our task therefore  is     to find
the discrete Lax operators  $U^\pm_j $ and  $\ F^d_0,  $ which in one hand  would  satisfy the
QYBE (\ref{lqybe}) with $R^{trig}
$-matrix (\ref{R})
 and on the other hand would recover the  DSG field model, we have started
with.
 That is the construction
of  the exact Lax operator  solutions of the QYBE: \   $U^\pm_j $
and $ F^d_0 $
would be such that
 they would reduce in the continuum limit (lattice constant $\Delta \to 0
$) to the field Lax operator of the SG model $ U(x, \xi)$  (\ref{UV})
and the BT matrix $  F^0_0$ (\ref{F0}),  respectively. For such a construction we turn to the
general scheme of   \c{kunprl}, where
 it was  shown   that any Lax operator
  satisfying  the QYBE (\ref{lqybe})
with  $R^{trig} $-matrix (\ref{R}) can be obtained  as  a particular realization of
  the  ancestor Lax operator
 \be
L_{anc}^{trig}(\xi) = \left( \begin{array}{c}
  \xi\hat c_1^{(+)} e^{i \al s^3}+ \xi^{-1}{\hat c_1^{(-)}}  e^{-i \al s^3}\qquad \ \
2 \sin \al  s_q^{(-)}   \\
    \quad
2 \sin \al  s_q^{(+)}    \qquad \ \  \xi{\hat c_2^{(+)}}e^{-i \al s^3}+
\xi^{-1}{\hat c_2^{(-)}}e^{i \al s^3}
          \end{array}   \right), \quad
          \ll{aL} \ee
% $U^\pm(s_q^\mp,c^\pm_a,s^3)$,
 %. We recall that for quantum integrability
with the quantum spin  operators  generating
   a generalized     quantum algebra
\be [ s_q^ {(+)}, s_q^{(-)} ] =
 \left (\hat  M^{(+)} \sin (2 \al s^3) -i
\hat  M^{(-)}  \cos
( 2 \al s^3  ) \right){1 \over \sin \al}, \quad
[s^3,s_q^{(\pm )}] = \pm s_q^{(\pm )} , \ \ [ \hat  M^{(\pm )}, \cdot]=0.
 \ll{ancAlg} \ee
Here the deforming  operators
$\hat  M^{(\pm )}=
(\hat c^{(+)}_1\hat c^{(-)}_2 \pm \hat  c^{(-)}_1\hat c^{(+)}_2 )$
are expressed through  $\hat c^{(\pm )}_a, \ a=1,2$, which  are mutually commuting and
 central (superscripts ${(\pm)} $ here  are obviously different
from    the  field labels  $\pm $ used in the DSG).

We intend to construct the quantum Lax operators
$U^\pm_j$  and   $ F^d_0$   as a solution of QYBE in a unified way from the
same (\ref{aL}).
Note that a reduction as
$   \hat c^{(\pm )}_a =\mp i \Delta , a=1,2 ,$
takes  (\ref{ancAlg}) to  $su_q(2)$ algebra
\be [ s_q^ {(+)}, s_q^{(-)} ] =2 \Delta ^2
 \frac { \sin (2 \al s^3)}{ \sin \al}, \quad
[s^3,s_q^{(\pm )}] = \pm s_q^{(\pm )} , \ \ q=e^{i \alpha}
 \ll{su2q} \ee
which can be realized
 in canonical variables  $[u^\pm_j,p^\pm_k]=i \delta_{jk} $ as
\be
s^3=\frac {u^\pm} {2}, \ s_q^ {(+)}(u^\pm, p^\pm)= e^{-2i p^\pm}g(u^\pm, \Delta),
 \ s_q^ {(-)}=(s_q^
{(+)})^\dagger,
\ll{lsg}\ee
{where} \ \
%\be
$g(u, \Delta)=\left (1 +\Delta ^2\cos \alpha (u + 1)\right)^{\ha} \frac {1} {\sin
\alpha}.$

%\ll{gu}\ee
Inserting the generators of $su_q(2)$ realized as (\ref{lsg}) in
  (\ref{aL}) we recover  the
 quantum Lax operators
: \be
U^{\pm}_j = \left( \begin{array}{c}
  \sin \al (\lambda+ \frac {u^\pm_j} {2})\qquad \ \
 \sin \al  s_q^{(-)}(u^\pm_j, p^\pm_j)   \\
 \sin \al  s_q^{(+)}(u^\pm_j, p^\pm_j)    \qquad \ \
  \sin \al (\lambda- \frac {u^\pm
 _j} {2})
          \end{array}   \right), \quad
\xi=  e^{i \al \la }        \ll{dsgL} \ee
 for the fields $u^\pm_j $ on the lattice
in conformity with the SG model  \c{qsgL}.

% \be L^{SG\pm}_j(\lambda) = \left( \begin{array}{c}
%  \Delta \sin  {\alpha}  ( \lambda+ \frac { u_j^\pm} {2})
%\qquad \ \
%\sin \al  g(u_j^\pm, \Delta)e^{ip_j^\pm}  \\   \quad
% \sin \al  e^{-i p_j^\pm}g(u_j^\pm, \Delta)    \qquad \ \
%  \Delta \sin   ( \lambda- \frac { u_j^\pm} {2})
%          \end{array}   \right), \quad
%          \lambda=e^{i \la}, \ll{dsgL} \ee
%reproducing  the known
%arranged  at sites $ \ j=\pm 1, \ldots \pm N  $ along both sides of
% the defect point and exactly satisfying QYBE.

 For   constructing  the discrete  BT operator  $ F^d_0(\xi, u^\pm_0, p^\pm_0)$
again   from
 (\ref{aL}),
  we choose the reduction
$   \hat c^{(+)}_a = 1, \  \ \hat c^{(-)}_a = 0,\ a=1,2 $, giving  $\hat
M^{(\pm )}=0$ and
reducing  (\ref{ancAlg}) to a simpler  algebra
\be [ s_q^ {(+)}, s_q^{(-)} ] =0,
 \quad
e^{i\al s^3}s_q^{(\pm )} =e^{ \pm i \al }s_q^{(\pm )}e^{i\al s^3}
 .\ll{alg0}\ee Fortunately, we can find
 a  consistent realization of the algebraic relations involving both the  canonical fields $
( u^\pm_0, p^\pm_0)$
  in the form
\be
e^{i \alpha s^3}= e_-, \ s_q^ {(+)}= (s_q^
{(-)})^\dagger= a  e_+ \ P_-^{-1},
\ \
  \mbox{where} \ \ e_\pm= e^{i {\al \over 4} (u^+_0\pm  u^-_0)},
P_-=e^{i {2  } (p^+_0-  p^-_0)}. \ll{rfd}\ee
 With commutation
relations $[e_+,P_-]=0, \ e_-P_-= e^{-i \alpha} P_-e_-  $, it is instructive
to check that the generators (\ref{rfd}) satisfy the algebra (\ref{alg0}).
 Therefore the Lax operator (\ref{aL})
reduces  finally to the explicit form
\be
F^d_0(\xi, u^\pm_0, p^\pm_0) = P_-^{\frac 1 2 \sigma_3} F^0_0(\xi,  u^\pm_0)
  P_-^{-\frac 1 2 \sigma_ 3}, \ \ F^0(\xi,u^+, u^-) = e^{- {i \alpha \over 4}\sigma_3   u^-} M(\xi, a)
e^{ {i \alpha \over 4}\sigma_ 3  u^+},
%\left( \begin{array}{c}\lambda  e_- \qquad \ \ a (e_+)^{-1} P_-
% \\   \quad  -a (e_+) P_-^{-1} \qquad \ \ \lambda   (e_-)^{-1}
%          \end{array}   \right),
  \ll{qdsgF} \ee
connecting remarkably to the BT operator  $ F^0_0(\xi,  u^\pm_0 )$ (\ref{F0}) for the DSG
model.
Note that  both the above discrete  Lax operators obtained
   as
realizations of the quantum integrable
 $ L$-operator (\ref{aL}), by construction must
satisfy the QYBE  exactly      with the $R^{trig}$-matrix. Consequently,
 (\ref{qdsgT}) with (\ref{qT+-})  represent    a  quantum
integrable  discrete DSG model.
% in analogy with   the solution of the
% standard SG model \cite{qsg}.
% However before that we have to
% ensure  that  the  integrable
% lattice model at the continuum limit would yield   the
%SG  field model with defect that we have constructed and investigated  above.
%check25
\subsection{Exact classical integrability }
As  the quantum Lax operator should satisfy
the QYBE for its quantum integrability, the corresponding classical Lax operator consequently
should satisfy its classical analog
 the classical YBE
(CYBE)\c{csg}
\be
\{ L_j(\xi), \otimes L_j (\eta)\}_{PB}=\de _{jk}[r(\xi,\eta),
L_j(\xi) \otimes L_k (\eta)  ]. \ll{cybe}\ee
 Note that at the classical
limit $ \bar h \to 0$, the quantum commutator should   reduce to the
Poisson bracket (PB) and the operator elements of the quantum Lax
operator would become just functions, with    the form of the Lax
operators (\ref{dsgL}, \ref{qdsgF}) remaining the same. Only the
normal ordering needed in the quantum case should  be ignored now.
However since the $ R$-matrix provides the structure constant for
the commutation relations, for transition to the  classical limit
we have to scale the parameter  $\al $  in the  $R$-matrix (\ref{R}) as
$\al \to \bar h \al $. This  therefore  defines the classical
limit as $\al \to 0$ in all the elements of the quantum $R$-matrix
  reducing it to the classical $r$-matix: $ \frac 1 {\sin \la
} R (\la, \al) = I+ \al r(\la) +o(\al ), \  r(\la)= \frac 1 {\sin \la
}\partial_\al R |_{\al
=0}$ as
\bea
& &r^{11}_{11}=r^{22}_{22}=a_0(\la ), \ r^{11}_{22}=r^{22}_{11}=0, \
r^{12}_{21}=r^{21}_{12}= c_0 (\la) \nonumber \\
\mbox {where } & & a_0(\la )=\cot (\la ),
\ c_0(\la )=\frac 1 {\sin \la
},
\ll{r0}\eea
while the QYBE   (\ref{lqybe}) reduces   to the
CYBE (\ref{cybe}).
 The underlying   algebraic relations are turned into PB relations at this classical  limit, which is
achieved by taking
 $\al \to 0$ in  (\ref{ancAlg}), only in the terms   that
come  from the $ R$-matrix.
As a result the classical limit of  (\ref{ancAlg}) is given by
\be \{ s_q^ {(+)}, s_q^{(-)} \}_{PB} =
 \left (\hat  M^{(+)} \sin (2 \al s^3) -i
\hat  M^{(-)}  \cos
( 2 \al s^3  ) \right){1 \over \sin ^2 \al}, \quad
\{e^{i \al s^3},s_q^{(\pm )}\}_{PB} =  \pm i  e^{i \al s^3} s_q^{(\pm)} ,
 \ll{PBanc} \ee
Therefore similar to the quantum case,
any realization of the classical limit of the $L$-operator (\ref{aL}),
 with its elements satisfying the PB relations
(\ref{PBanc}) must be a solution of the CYBE, by construction.
%CHECK251
 Consequently, since  $ U^\pm_j $  (\ref{dsgL}) and  $ F^d_0 $   (
\ref{qdsgF}), as shown above, are indeed  realizations of
 (\re{aL}), their classical limit  must satisfy the CYBE (\ref{cybe}) exactly, proving  the
classical integrability of the DSG model. Alternatively one  can
 check this statement  through direct verification. For example
at $\hat  M^{(+)}=2\De ^2, \hat  M^{(-)}=0 $, i.e. at the $su_q(2) $ limit
the PB relations (\ref{PBanc}) reduce to
\be \{ s_q^ {(+)}, s_q^{(-)} \}_{PB} =
 2 \De ^2{ \sin (2 \al s^3)  \over \sin ^2 \al}, \quad
\{e^{i \al s^3},s_q^{(\pm )}\}_{PB} =  \pm i  e^{i \al s^3} s_q^{(\pm)} ,
 \ll{PBsuq} \ee
which is evidently satisfied by the realization of the generators in the
canonical fields  $\{u^\pm_j,p^\pm_k\}_{PB}= \delta_{jk} $ as
\be
s^3=\frac {u^\pm} {2}, \ s_q^ {(\pm)}(u^\pm, p^\pm)= \frac 1  {\sin
\alpha} e^{\mp2i p^\pm}(1+\De ^2 \cos \alpha u ) ^{\ha}
\ll{csgr}\ee
using the definition $\{ f, g\}_{PB} =\frac {\partial f} {\partial u}\frac {\partial
g} {\partial p}-\frac {\partial f} {\partial p}\frac {\partial
g} {\partial u}  .$
In a similar way we can check the validity of the required PB relations for
 the elements of   $ F^d_0 $   (\ref{qdsgF}), obtained
from (\ref{PBanc})  at $\hat  M^{(\pm)}=0 $:
\be \{ s_q^ {(+)}, s_q^{(-)} \}_{PB} =0,
 \quad
\{e^{i\al s^3}, s_q^{(\pm )} \}_{PB} = \pm i e^{i\al s^3}s_q^{(\pm )} .
 \ll{pbF} \ee
Using the classical analog of  (\ref{rfd}) expressed  in the same form
without normal ordering:
$ e^{i \alpha s^3}= e_-, \ s_q^ {(\pm)}= a  e_+ \ P_-^{\mp 1},
\ $ and the PB relations like $ \{e_+,P_-\}_{PB}=0, \ \{e_-,P_-^{\mp 1}\}=
\mp i e_-,P_-^{\mp 1}$ we  verify the relations (\ref{pbF}), which
guarantees   that  $ F^d_0 $  satisfies the CYBE (\ref{cybe}).
The solution of the local CYBE (\ref{cybe}) leads also to that for the
global CYBE for $T(\xi)=\prod_jL_j(\xi) $:
\be \{ T(\xi), \otimes T (\eta)\}_{PB}=\de _{jk}[r(\xi,\eta),
T(\xi) \otimes T (\eta)  ],
 \ll{gcybe}\ee
 from where in exact analogy with the SG model \c{FT} one can extract the
action-angle variable corresponding to the continuum as well as the discrete
spectrum, proving the
complete classical integrability of the  DSG model.
 This also reveals an  intriguing fact that
 the local differences between the SG and the DSG models  seem to become irrelevant
at the global level (\ref{gcybe}).

It is remarkable  that at the continuum limit $\Delta \to 0$,  when
the discrete Lax operator $F^d_0 \to F^0_0 $
recovers the BT matrix  (\ref{F0}), the dependence of the canonical momentum
$P_- $ drops out completely, creating a paradoxical situation that would result
 a trivial PB for the $F^0 $ and  naturally  not satisfying  the CYBE (\ref{cybe}).
 Therefore for showing the validity of the CYBE for the field
model one has to be careful and
 should take first the PB in the corresponding exact discretized model
  and then  perform the continuum limit.
Similar is true also for the Lax operators $ U^\pm_j  $.

Our another important result is a discretized BT relation that we can derive
connecting the discrete sine-Gordon Lax operators $U^\pm_j $ (\ref{dsgL})
 in the form of
a discrete gauge transformation
\be
U^+_j(\xi,u^+_j)=F^d_{j+1}(\xi, u^\pm_{j+1}, p^\pm_{j+1})U^-_j(\xi,u^-_j)
(F^d_{j}(\xi, u^\pm_{j}, p^\pm_{j}))^{-1},
  \ll{dBT} \ee
through the gauge matrix $F^d_{j}(\xi, u^\pm_{j}, p^\pm_{j}) $ (\ref{qdsgF}).
Note that we assume this  discretized BT to be  valid for arbitrary  $j$.
At the continuum limit the  discrete gauge transformation (\ref{dBT}) should recover the known
gauge relation (\ref{btgt}) connecting the field Lax operators $U^\pm(u^\pm) $.
 Similarly comparing the elements of the matrix relation
(\ref{dBT}) we can obtain the bridging relation between the discrete variables
$u^+_j $ and $u^-_j $ of the discretized DSG model at any site $j$,
 involving also  the variables $u^\pm_{j+1},p^\pm_j, p^\pm_{j+1} $. This
 discrete  BT relation would yield  at the continuum  limit $
\De \to 0$,  the standard BT relation between the fields $u^\pm(x) $ (\ref{btgx}).
\subsection  { Algebraic Bethe ansatz}
  Following the formulation of quantum SG model \c{qsg}
 we can apply the Algebraic Bethe ansatz method   to the lattice regularized
quantum DSG  constructed above and solve in principle
 its eigenvalue problem exactly.
Recall that
the aim of the algebraic Bethe ansatz
 is to solve exactly the eigenvalue problem of
$\tau (\xi)=trace T(\xi ), \ \ T(\xi )=\prod_j L_j(\xi) $, generating all conserved
operators including the Hamiltonian, in the form: $\tau(\xi)|n >=\Lambda(\xi)|n >
$,
 with the eigenstates $ |n > $  defined as
  $|n >= |\xi _1, \ldots, \xi_ n>= \prod_s^n B(\xi _s)|0>$.
 $T_{12}(\xi) =B(\xi )$ acts as  {creation} operator, while
$T_{21}(\xi) =C(\xi )$  as  { destruction} operator annihilating
the pseudovacuum: $C(\xi
)|0>=0. $
A crucial step in this formalism is to construct the
pseudovacuum state $|0> $, which  is  achieved
   for the SG model
 by combining the actions of  consecutive
pair of Lax operators: $U_jU_{j+1}|0> $ \c{qsg}.
Repeating this procedure
we can construct    the  pseudovacuum
 $|0>^\pm=\prod_{j=\pm 1}^{\pm \frac N 2 }|\Omega_j^{(2)}> $ for the  quantum DSG model, yielding $C^\pm(\lambda)|0>^\pm=0
$,
for   all sites  except the defect
point at $j=0 $.

%consecutive sites $U^\pm_{j+1}U^\pm_{j} $ are acted upon a
%pseudovacuum to construct it in the form $ f(q_1,q_2)=(1+\Delta ^2
%g(q_1,q_2))\delta (q_1-q_2-\pi+\al)$, where is known solution
%\c{qsg}.
 However the single  defect point would play
a nontrivial role, since after crossing this point, say from the left the
pseudovacuum property:  \be
T^{-}(\xi)|0>^- = \left( \begin{array}{c}
  A^-(\xi)|0>^- \qquad \ \
 |1>^-   \\
 0   \qquad \ \ \ \
  A^{-*}(\xi)|0>^-
          \end{array}   \right),       \ll{vsgL} \ee
 would be lost due to nontriangular matrix form of
$F^d_0|\Omega _0> $. Instead of annihilating the local vacuum, as needed by
 the lower left corner operator element  of the matrix
$F^d_0|\Omega _0> $,
the defect  at site $j=0 $  would turn it to a state $O|\Omega _0> , \
O=-a e^{i ({2  } (p^+_0-  p^-_0)+ \frac
\al  4  (u^-_0+u^+_0))} $,  creating at the same time its conjugate
state    $ -O^\dagger|\Omega _0>$ at the upper right corner.
 This is expected to lead to
   the
creation/annihilation of quantum states by the defect point similar to
that with  classical solitons as we have
observed  exploiting the BT in sect. 4.
Perhaps one should explore the possibility of using a quantum extension of
the BT \c{qBT} to mimic the successful classical approach, using the
relation like (\ref{dBT}) at the quantum level.
 These tricky points however need careful
and separate analysis and should be dealt with  elsewhere.

\subsection { Continuum limit}
 It is crucial to check that the  classical and quantum integrable discrete DSG
model
 we constructed and solved above
  would yield
 the same  DSG field model we have started with,  at the
  continuum limit: $\Delta \to 0 $. Note that
at this limit
  the  canonical variables go to
canonical fields:  $u^\pm_j \to u^\pm(x), p^\pm_j \to \Delta p^\pm(x)  $,
 with
$[u^\pm(x), p^\pm(y)]= i\delta(x-y) $.   Therefore for extracting the limit
we have to scale $  p^\pm_j $  giving $ e^{ {i }
p^\pm_j } \approx 1+ i\Delta p^\pm(x), $
in both the discrete Lax operators $U^\pm_{j} $ and $F^d_0 $.

 Observing further  that {\it rotated} operator
$ \sigma_1U^{SG\pm}_j(\lambda) $ is also a solution of QYBE  due to the
symmetry of the $R$-matrix:
 $[R^{trig},\sigma_a \otimes \sigma_a]=0, a=1,2,3 $ , we expand the   Lax operator (\ref{dsgL})
in powers of $
\Delta $ at all  sites $ j$
 (except at the defect point) as
\be
 \sigma^1 U^{\pm}_j
= (1+ \int ^{x+ \frac \Delta 2 }_
{x- \frac \Delta 2 }U^{\pm}(x') dx')+O(\Delta ^2)
 = (1 +\Delta U^{\pm}(x))+O(\Delta ^2) , \ll{l}\ee
with the field  operator $U^{\pm}(x) $ recovering
exactly  the    Lax operator (\ref{UV})
 of the SG field model. Note that we have put the deforming parameter $
\alpha =1$ in all expressions related to the continuum model, for
simplicity.
 Thus  we recover
 at the continuum limit, the DSG field model at all
points except at $x=0 $.

% where  \be
%U^{SG\pm}_j(x) = \frac 1 4i \left( \begin{array}{c}-i p^\pm \qquad \ \
%\sin  {\alpha}  ( \lambda+ \frac { u^\pm (x)} {2}) \\
%\sin  {\alpha}  ( \lambda- \frac { u^\pm (x)} {2}) \\ \qquad \ \ -i p^\pm
%          \end{array}   \right).  \ll{sgL} \ee
 Performing   the same continuum  limit at  the  defect point
 $ j=0$ ,  we get on the other hand from
(\ref{qdsgF})
:
\be
F^d_0(\xi,u^+_0,p^+_0; u^-_0,p^-_0) \to  F^0_0(\xi,u^+(0), u^-(0)) +
i \Delta F^{'}, \ \ F^{'}=(p^+_0-p^-_0)[\sigma_3 , F^0_0(\xi,u^+(0), u^-(0))],
\ll{fx}\ee
 clearly giving  $F^0 $ at $\Delta \to 0 $, i.e. it recovers the same
BT matrix (\ref{F0}) at the defect point, meeting the essential requirement.
Therefore  collecting all nontrivial terms
  we get finally  the continuum limit of the lattice regularized model  (\ref{qdsgT})
as
\be
T(\lambda)=\left(e^{ \int ^{+ \infty}_{0}
U^{+}(\lambda, x') dx'}\right)\  F^0_0(\lambda,u^+(0), u^-(0))
\ \left(e^{ \int _{- \infty}^{0}
U^{-}(\lambda, x') dx'}\right)
\ll {Tcsg} \ee
 yielding the  original  DSG field model.

Finally  in the continuum limit using  (\ref{l}) and  the expansion like
$F^d_{j+1}\to  F^0 +\De (F^0_{x}+iF^{'}), \ u^\pm_{j+1} \to u^\pm+\De u^\pm_{x},
\ p^\pm_{j+1} \to \De (p^\pm+\De p^\pm_{x}) $ etc. we can show directly that
the discrete BT relation (\ref{dBT}) goes to the field BT as gauge-transformation
between  $U^\pm(u^\pm) $, while the relation between its matrix elements
 connecting   $u^\pm_j $ recovers the bridging relation
(\ref{btgx}).

We stress again that for proving the classical and quantum integrability of
the field model   (\ref{Tcsg}) one has to go to its proper lattice
regularized version  (\ref{qdsgT}) and check the validity of the CYBE and
QYBE before taking back the continuum limit.

%%%%%%%55beore conclusion%%%%%%%%%%%%%%%%%%%%%%%%%%%%%%%%%%

\section {Concluding remarks}
We have proved here the  classical and quantum integrability
of the sine-Gordon model with a defect  by finding the exact
solution of the related quantum as well as classical Yang-Baxter
equation through  integrable discretization of the model. This enables one to
solve the quantum eigenvalue problem by  algebraic Bethe
ansatz and  the  classical model for the action-angle
variables.

Combining Matrix B\"acklund transformation with
  the matrix Riccati equation approach we  have extended the existing formalism
 and found  all higher conserved quantities for the defect sine-Gordon  model in a
systematic way, with explicit forms of the defect contribution. In the
simplest case this gives the momentum and Hamiltonian of the DSG
model found earlier. Deviating from the earlier studies we have used the
{\it unfrozen} BT, which  can  produce
 intriguing effect like creation or annihilation of solitons by the defect
point, apart from the
preservation of soliton with phase shift as predicted earlier.
We  have also found the  constraints showing  exactly how 
 the creation/annihilation or preservation of solitons 
 depend
on the boundary conditions of the field, 
 in the framework of the defect sine-Gordon  model. This result   should be of crucial
importance   
 in the possible experimental  detection of such unusual
events. 
A pertinent question arises here regarding the obvious
 violation of topological
charge in this SG model with a defect due to possible nonconservation of soliton number.
 It should be noted however that the
topological charge arises in the SG model as a degree of mapping from $S^1 \to
S^1$, while  a defect in the coordinate-axis or a discontinuity point (like
a puncture in the sphere) as occurs in the DSG, can not be mapped into a smooth
sphere or $S^1 $, violating thus the concept of the topological charge
itself. Therefore in the DSG the solitons seem to be  no longer topological and
hence
their number may change.
  The formation of nontopological semi-fluxons
\c{msgni1}  therefore  may also be possible to explain  in the framework of the DSG
model \c{kunfutur}.

After completion of  our work, very recently  an important paper
has appeared in the arXiv \c{def07}, where  systematic studies were
made  using  the Lax pair  formalism of integrable systems in the
line of the present investigation, and consequently infinite set
of conserved quantities were obtained for a whole class of defect
models e.g. SG, NLS, KdV, mKdV, Liouville, DNLS etc. ,
belonging to the   AKNS and the KN spectral problems and
having  a defect at a single point, which in principle could be
extended to multiple defect  \c{def07}. Similar to our approach
the conserved quantities are found through Riccati equation, though
in contrast to the Matrix  Riccati equation used here  this equation is a scalar one.  In
the same work the importance of   establishing the  complete
integrability of the defect models, at the classical and the
quantum level, through the Yang-Baxter equation as well as  the
necessity  of discretization of the model for achieving this goal
are emphasized. Interestingly both these  agenda in the  wish-list
were addressed and solved rigorously in the present paper.
%%%%%%%%%%%%%%%%%%%%%%%%%%%%%%%%%%%%%%%%%%%%%%%%%5

Extension of the exact  result of the quantum and classical DSG
model through the Yang-Baxter equation presented here,  to other models like NLS, DNLS
etc. treated in \c{def07} would be an interesting problem.

\end{document}